# MARS: Multi-macro Architecture SRAM CIM-Based Accelerator with Co-designed Compressed Neural Networks

Syuan-Hao Sie, Jye-Luen Lee, Yi-Ren Chen, Chih-Cheng Lu, Chih-Cheng Hsieh, Meng-Fan Chang, *Fellow*, *IEEE* and Kea-Tiong Tang, *Senior Member*, *IEEE*

**Abstract** Convolutional neural networks (CNNs) play a key role in deep learning applications. However, the large storage overheads and the substantial computation cost of CNNs are problematic in hardware accelerators. Computing-in-memory (CIM) architecture has demonstrated great potential to effectively compute large-scale matrix-vector multiplication. However, the intensive multiply and accumulation (MAC) operations executed at the crossbar array and the limited capacity of CIM macros remain bottlenecks for further improvement of energy efficiency and throughput. To reduce computation costs, network pruning and quantization are two widely studied compression methods to shrink the model size. However, most of the model compression algorithms can only be implemented in digital-based CNN accelerators. For implementation in a static random access memory (SRAM) CIM–based accelerator, the model compression algorithm must consider the hardware limitations of CIM macros, such as the number of word lines and bit lines that can be turned on at the same time, as well as how to map the weight to the SRAM CIM macro. In this study, a software and hardware co-design approach is proposed to design an SRAM CIM–based CNN accelerator and an SRAM CIM–aware model compression algorithm. To lessen the high-precision MAC required by batch normalization (BN), a quantization algorithm that can fuse BN into the weights is proposed. Furthermore, to reduce the number of network parameters, a sparsity algorithm that considers a CIM architecture is proposed. Last, MARS, a CIM-based CNN accelerator that can utilize multiple SRAM CIM macros as processing units and support a sparsity neural network, is proposed.

## I. Introduction

DEEP neural networks (DNNs) have demonstrated outstanding performance in various domains because of their highly flexible parametric properties. Among these domains, computer vision has had the most progress because many different convolutional neural network (CNN) architectures have been designed. VGG [1] uses multiple 3 × 3 kernels instead of one large kernel to reduce network parameters and increase depth. ResNet [2] uses the concept of residuals to return the deep gradient value to the shallow layer in order to avoid gradient vanish. Ioffe, et al. [3] designed batch normalization (BN) to reduce internal covariate shift. MobileNet [4] uses depthwise convolution and pointwise convolution to reduce the number of parameters in networks.

However, the high computational complexity and high energy consumption of CNNs hamper their application in mobile devices. In hardware, various CNN accelerators have been proposed to address computing needs. Eyeriss [5] utilized row-stationary data flow to accelerate convolutional operations. QUEST [6] realized data exchange between processing units (PEs) and static random access memory (SRAM) by using three-dimensional stacking. Sticker [7] supported multiple-sparsity and dense convolution. However, the aforementioned CNN accelerators are still based on the Von Neumann architecture, which requires substantial amounts of energy to transfer the massive amounts of data between memory and PEs. As Moore's law saturates [8], reducing the energy consumption caused by memory access becomes increasingly difficult. To break the Von Neumann bottleneck, computing-in-memory (CIM) has been investigated. The CIM structure enables the reduction of large amounts of intermediate data transfer, highly parallel computing, and low standby power consumption. In addition, CIM is considered suitable in CNNs that are required to calculate large amounts of matrix-vector multiplications. Non-volatile CIM, such as resistive random access memory (ReRAM), exhibited high computational efficiency in analog

This paper submitted at 2020-07-xx. This work was supported by the Ministry of Science and Technology, Taiwan, under contract no. MOST 109-2218-E-007-019 and MOST 108-2262-8-007-017.

SyuanHao Sie, Jye-Luen Lee, Yi-Ren Chen , Chih-Cheng Hsieh, Meng-Fan Chang, and Kea-Tiong Tang are with National Tsing Hua University (NTHU), Hsinchu 30013, Taiwan. (email: kttang@ee.nthu.edu.tw).

Chih-Cheng Lu is with Information and Communication Labs, Industrial Technology Research Institute, Chutung 31030, Taiwan.

dot products [9]. However, challenges regarding large-scale manufacturing and the limited durability of memristors have affected the development of ReRAM-based CIM. Therefore, from the perspective of technology availability, some researchers have started to develop SRAM-based CIM (SRAM CIM) for artificial intelligence tasks [10]–[18].

In software, to reduce storage and computation costs, different model compression algorithms have been proposed, particularly the sparsity algorithm and the quantization algorithm. The sparsity algorithm limits the parameters during training by designing a loss function; in inference, the parameters under the threshold are set to zero; this processing is called pruning. Depending on restrictions, pruning type can be divided into filter-wise [19], channel-wise [20], shape-wise [21], and fine-grained pruning [22] (Fig. 1). In the quantization algorithm, the input and weight bit-width is limited to reduce the computational complexity by using different types of quantizers, including binary [23], ternary [24], uniform [25]–[27], and non-uniform quantizers [28]–[30].

Despite extensive research on software and hardware, only a few of these software methods can be applied in hardware; many prior works have ignored the architecture and operating mechanisms of CNN accelerators. Considering the aforementioned factors, in this work, a multi-core architecture SRAM CIM-based accelerator called "MARS" and model compression algorithm according to a SRAM CIM-based architecture are proposed. The contribution of this work is as follows:

- MARS, a SRAM CIM-based accelerator with eight SRAM CIM macros, is proposed. The accelerator can support a sparse CNN with high throughput and energy efficiency.
- A sparsity algorithm that considers the word lines (WLs) and bit lines (BLs) of SRAM CIM is proposed that can achieve a high compression rate with low index storage requirements.
- A quantization algorithm with BN fusion is proposed to lessen the high-precision multiply and accumulation (MAC) operations required of the hardware.

The remainder of this paper is organized as follows. Section II introduces the background of SRAM CIM, SRAM CIM–based accelerators, and model compression. Section III introduces the architecture and the data flow of MARS. Section IV describes the model compression algorithm. Section V presents the experimental results, and Section VI concludes this paper.

## II. Background

### A. SRAM CIM Macro

In computing systems with von Neumann architecture, breaking through the "memory wall" bottleneck is difficult, particularly when a large amount of data movement takes place between the PEs and memories. CNN training and inference typically require modifying data and parameters frequently.

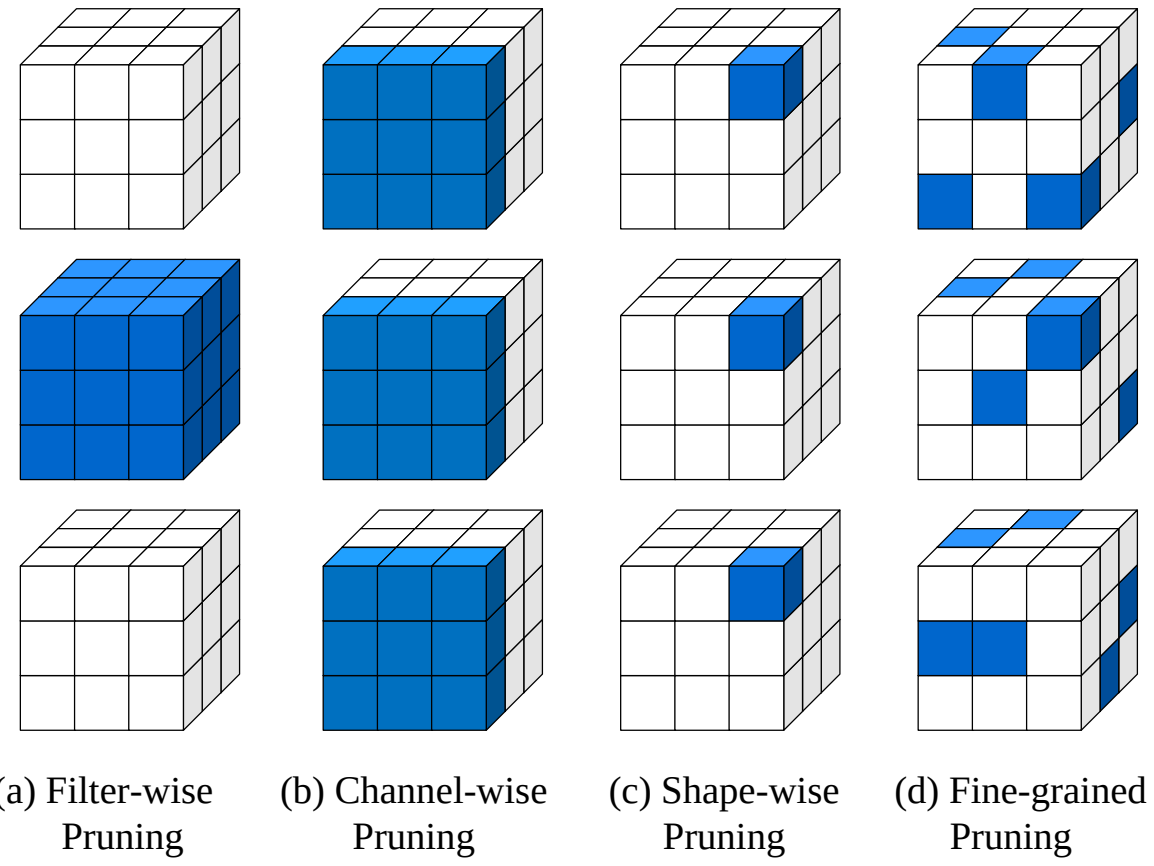

Fig. 1 Different pruning types. (a) Filter-wise and (b) channel-wise pruning are hardware friendly but accuracy drops considerably. Although (c) shape-wise and (d) fine-grained pruning can maintain greater accuracy, special hardware architecture is required.

CIM has been widely researched because of its ability to reduce data movement and perform arithmetic operations while storing data in the same place. To address the requirements of different applications, many silicon-verified SRAM CIM macros have been proposed. Zhang et al. proposed a machine-learning classifier that was implemented in a 6T SRAM array [10]. Si et al. proposed a dual-split-control 6T SRAM CIM that can support a fully connected layer [11]. Biswas et al. proposed a 10T Conv-SRAM for binary weight neural networks [12]. Jiang et al. proposed an XNOR-SRAM for binary/ternary DNNs [13]. Dong et al. proposed a 4+2T SRAM macro for embedded searching and CIM applications [14]. Hossein et al. proposed a charge-domain in-memory-computing accelerator for binarized CNN [16]. These CIM macro works have demonstrated various benefits of CIM in terms of functionality and improved energy efficiency. In the current work, the state-of-the-art 6T-SRAM chip [18] is adopted to design our hardware architecture, and the algorithm also followed the specifications of this chip.

### B. CIM-Based Accelerator

Many researchers have designed CIM-based accelerators according to the ideal CIM macro, particularly the ReRAM-based accelerator, because ReRAM has less area and high energy efficiency. ISAAC [31] was a well-known ReRAM-based accelerator that supported high-precision fixed-point arithmetic. In addition to the general hardware architecture, some researchers have attempted to design sparse PEs and pruning algorithms considering the properties of CIM macros. RECOM [32] was the first CIM-based accelerator to support sparse DNN processing. Through appropriate design of the corresponding circuit, the weights, which are all zero in the same rows or columns can be skipped. Furthermore, [33] proposed a sparse CNN-mapping algorithm based on k-means clustering to eliminate all-zero crossbars. However, in the aforementioned cases, it was assumed that the CIM macros were ideal. Few studies have designed a CIM macro for system-level architecture. The tape-out CIM macro is different from the ideal CIM macro in several aspects. First, the ideal CIM macro frequently has a large capacity, which minimizes the likelihood of reloading during calculation. Second, the ideal CIM can realize high-precision calculations, whereas the tape-out CIM can only implement low-precision calculations. Third, the ideal CIM macro can calculate all the weights in the same cycle, whereas the tape-out CIM macro must consider the number of analog-to-digital converter (ADC) and the variation caused by the BL current. Therefore, only a limited number of weights can be calculated in a single cycle. To resolve these problems, this work proposes a hardware and software co-design to compensate for the limitation of the CIM macro.

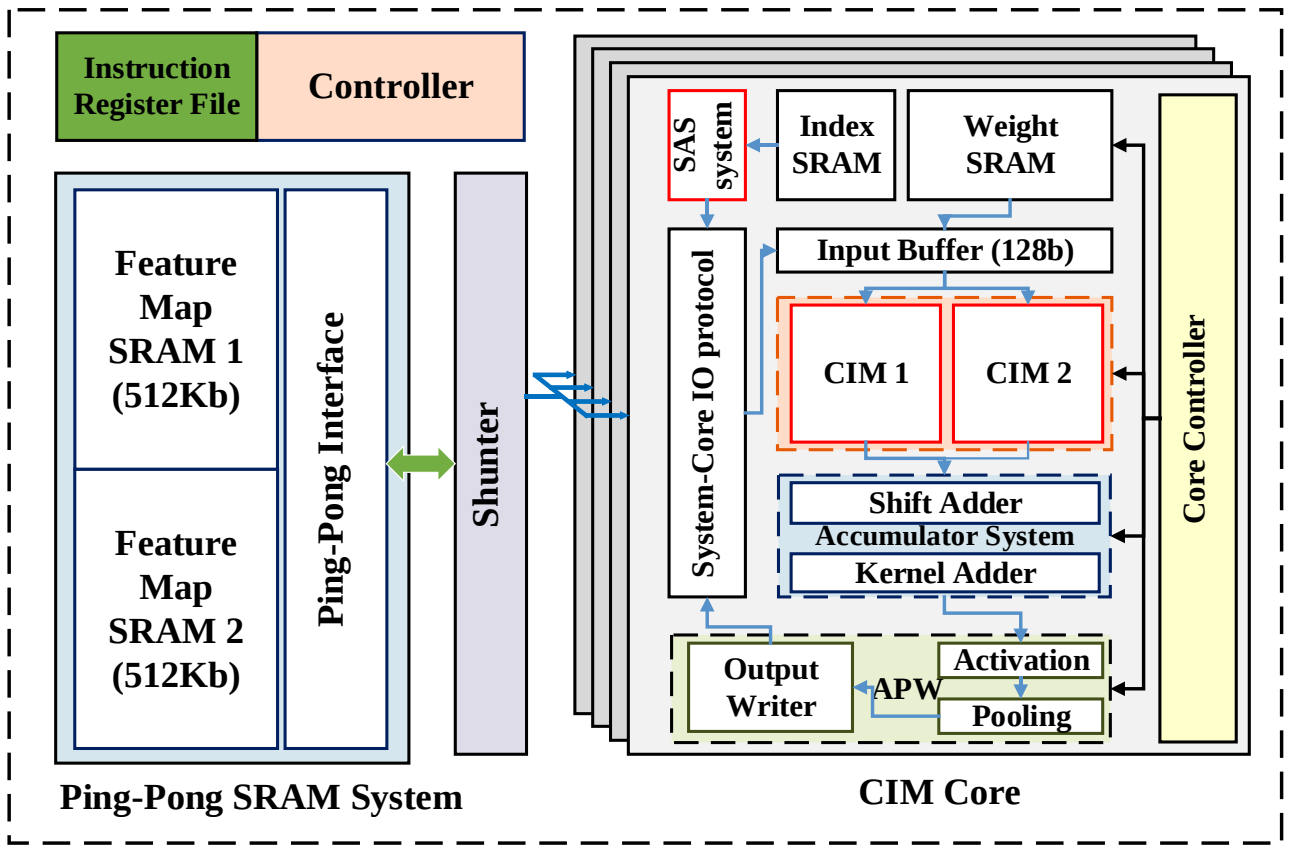

Fig. 2 Overall architecture of MARS.

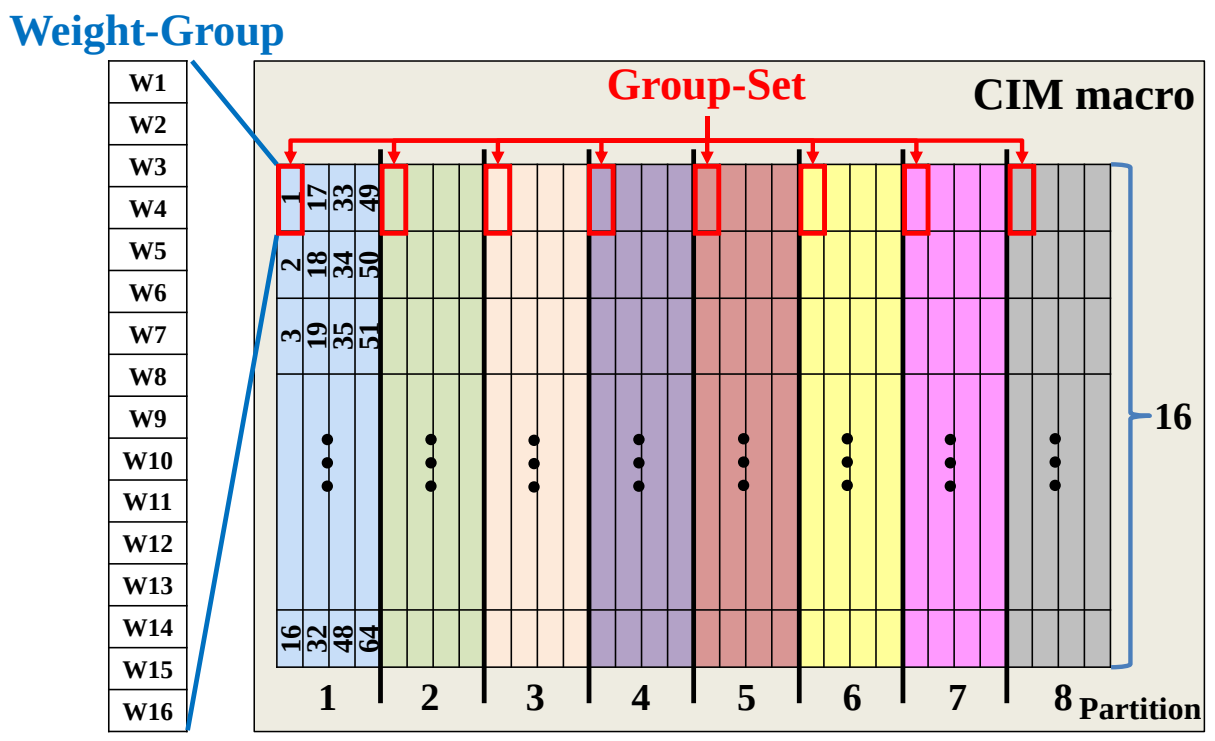

Fig. 3 The SRAM CIM macro contains 8 partitions, and each partition contains 64 weight-groups. Each grid represents one weight-group, and the 16 weight-groups in each partition at the same position is considered as a group-set.

### *C. Model Compression*

Conventionally, a CNN model is trained with floating point, and then is converted to fixed point to reduce memory overhead when deployed on hardware; this process is called model quantization. Model quantization can be divided into uniform quantization [25], [26] and non-uniform quantization [28]–[30]. Generally, non-uniform quantization enables better performance than uniform quantization, but uniform quantization has hardware-friendly properties that can be directly implemented on off-the-shelf accelerators with bit operation or integer-only arithmetic. BN [3] is a common method for improving model performance; however, implementing BN on an edge device requires the design of additional circuits to implement high precision MAC, increasing the area and complexity of the whole system. In this work, a new method is proposed to fuse BN into quantize-aware training. As a result, the model can be benefited by the advantage of BN without adding additional circuits in the accelerator.

Another common method used to reduce weights is model pruning. Since 2015, many studies on CNN model pruning have proposed shrinking the model size to reduce resource and energy consumption. Although filter-wise [19] and channel-wise pruning [20], as shown in Fig. 1, are hardware friendly and suitable for any hardware architecture, these two methods typically exhibit considerable accuracy loss and low compression rate. Therefore, several studies have focused on how to implement fine-grained pruning efficiently on hardware because of its higher compression rate [5], [34]–[35]. However, because of the limitations of the CIM architecture, the weights can only be skipped when all the values on a WL are zeros. The irregular pattern of fine-grained pruning is also inefficient for mapping on the CIM macro.

## III. Proposed MARS Accelerator

In this section, an SRAM CIM-based accelerator that adopts the SRAM CIM proposed in [18] as the main processing and memory unit is proposed. MARS has the ability to skip the zero weight calculation and compress the memory usage of weights, which further reduces the operation time in CNN convolution and data transfer between memory and CIM. The architecture of MARS can be mainly divided into two parts: the first part is the top-level system responsible for the whole system control, and the second part is the CIM core focusing on data processing and storage.

### *A. Architecture Overview*

1) Top-level system

The top level is composed of a controller, Ping-Pong SRAM system, and four CIM cores, as shown in Fig. 2. The controller extracts the instruction code stored in the instruction register file and sends the corresponding control signal to the whole system. To avoid the additional transfer of the feature map, which consumes energy and time, the Ping-Pong SRAM system is adopted. Two 512-Kbit feature map SRAMs (SRAM1,2) are used in the Ping-Pong SRAM system to store the input feature map (IFM) and the output feature map (OFM). These FM SRAMs can be accessed by the four CIM cores with the help of the shunter module. The two FM SRAMs can be switched to store either IFM or OFM through the instruction code assignation of every layer during the CNN model calculation.

2) CIM core

The CIM core mainly contains two SRAM CIM macros, weight SRAM, index SRAM, a sparsity address search (SAS) system, an accumulator, and an activation-pooling-write (APW) block. Each SRAM CIM macro can accommodate 8192 × 8 bits (64Kb). Because of CIM's highly parallel computing with favorable efficiency, the conventional digital PE is replaced with CIM macros. Taking advantage of SRAM CIM being both memory and a PE, the energy consumption of transfer weighted between the memory and the PE is minimized. Index SRAM stores the index code of nonzero weight positions, and the controller module controls Index SRAM to provide current layer's kernel sparsity information to the SAS module to generate the correct corresponding IFM address. Weight SRAM accommodates the weight of the current layer because CIM cannot store all weight at once; therefore, the CIM macro must reload new weights for each new layer. The result of the internal calculation of the CIM is accumulated by the shift accumulator and the kernel accumulator. The accumulated result is further processed by the APW block, and the calculated result is stored in the OFM SRAM.

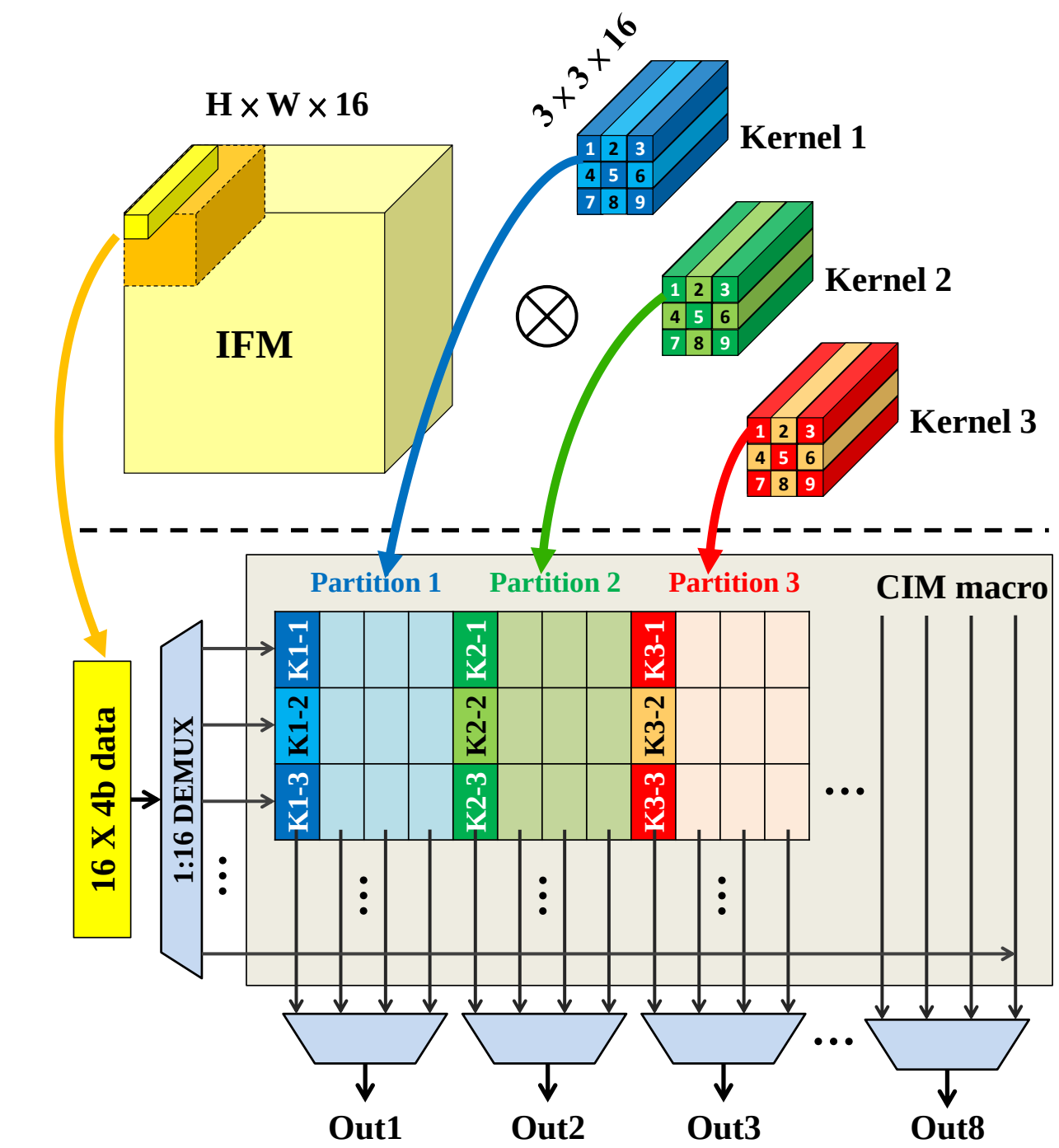

Fig. 4 CIM's highly parallel computing feature can be viewed as eight convolution operations at one time. This figure is an example of an IFM convolving with three 3x3x16 kernels and mapping these kernels into different CIM partitions.

## *B. Sparsity Mechanism*

### 1) SRAM CIM macro

The adopted SRAM CIM macro contains 8 partitions, and each partition can be divided into 64 groups of 16 weights. Hereafter, the 16 weights of a group are defined as a weight-group. When using an SRAM CIM macro for computing, each partition activates one weight-group at the same relative position through the control signal, as illustrated in Fig. 3. The 16 input data are shared in 8 partitions and perform the inner product with the activation of the weight-group at the same time, and then 8 results are generated in the next cycle. This inner product operation behavior is in accordance with the convolution calculation in CNN, as illustrated in Fig. 4. In this work, the two SRAM CIM macros are combined into one core sharing the same control signal and input to acquire higher parallel computation capability. By doing so, the 16 weight-group can be activated at the same time and perform 16-vector inner products of 16 kernels in one cycle. These 16 weight-group at the same relative position is defined as a group-set.

K1-10 : Zero weight-group(without color)   K1-1 : Nonzero weight-group(color)

| | | | |
|---|---|---|---|
| K1-16 | K2-16 | K3-16 | K4-16 |
| K1-15 | K2-15 | K3-15 | K4-15 |
| K1-14 | K2-14 | K3-14 | K4-14 |
| K1-13 | K2-13 | K3-13 | K4-13 |
| K1-12 | K2-12 | K3-12 | K4-12 |
| K1-11 | K2-11 | K3-11 | K4-11 |
| K1-10 | K2-10 | K3-10 | K4-10 |
| K1-9 | K2-9 | K3-9 | K4-9 |
| K1-8 | K2-8 | K3-8 | K4-8 |
| K1-7 | K2-7 | K3-7 | K4-7 |
| K1-6 | K2-6 | K3-6 | K4-6 |
| K1-5 | K2-5 | K3-5 | K4-5 |
| K1-4 | K2-4 | K3-4 | K4-4 |
| K1-3 | K2-3 | K3-3 | K4-3 |
| K1-2 | K2-2 | K3-2 | K4-2 |
| K1-1 | K2-1 | K3-1 | K4-1 |

(a) Traditional mapping

| | | | |
|---|---|---|---|
| K2-15 | K4-13 | K6-17 | K8-16 |
| K2-8 | K4-12 | K6-16 | K8-15 |
| K2-7 | K4-11 | K6-15 | K8-14 |
| K2-6 | K4-5 | K6-14 | K8-13 |
| K2-5 | K4-4 | K6-7 | K8-12 |
| K2-4 | K4-3 | K6-5 | K7-11 |
| K2-2 | K3-15 | K6-3 | K7-10 |
| K1-9 | K3-14 | K6-1 | K7-9 |
| K1-8 | K3-11 | K5-10 | K7-8 |
| K1-7 | K3-10 | K5-9 | K7-7 |
| K1-6 | K3-9 | K5-8 | K7-6 |
| K1-5 | K3-8 | K5-7 | K7-5 |
| K1-4 | K3-7 | K5-6 | K7-4 |
| K1-3 | K3-2 | K4-16 | K7-3 |
| K1-2 | K3-1 | K4-15 | K7-2 |
| K1-1 | K2-16 | K4-14 | K7-1 |

(b) Weight Sparsity mapping

Fig. 5 Weight mapping in one partition of the CIM. (a) Showing the mapping situation with traditional mapping method contains a lot of zero weight-group in CIM. (b) Showing that weight sparsity mapping method of MARS can accommodate more nonzero weight-group by not storing the zero weight-group

### 2) Weight Sparsity Mapping in CIM

In the conventional approach, all weights of a kernel must be stored in SRAM CIM macros; therefore, mapping weight to the SRAM CIM mostly follows the sequence according to kernel size, as illustrated in Fig. 5 (a). However, in the conventional approach, the controller of the SRAM CIM is straightforward to design, and system performance is restricted by the SRAM CIM macro. Typically, in a well-trained CNN model, numerous weights are zero, particularly in deep layers that consume substantial calculation effort. In such a case, repeatedly loading and computing zero weights in SRAM CIM is inefficient. Because zero multiplied by any number is zero, we can foresee the result without calculating the zero weight. Skipping the calculation of the zero weight can considerably reduce the operation time of the CNN. The CIM would also not be required to store zero weight because zero groups of weight in the SRAM CIM will not be accessed for computing. Therefore, in our design, if the weights in a group-set are all zero, we can skip the calculation of that group-set, and the 16 group-weights of that group-set will not be stored in the SRAM CIMs. Fig. 5 illustrates the difference between the conventional method and the proposed method of mapping the same kernels to the CIM. In contrast with Fig. 5 (a), Fig. 5 (b) reveals that when some weight sets are all zero, they are not stored in the CIM to avoid unnecessary calculations, and the SRAM can hold more data. This mechanism enables the SRAM CIM to only store meaningful nonzero weights. For that reason, SRAM CIM can accommodate network architectures that exceed its original capacity and further reduce the time required to reload weights to CIM. To maximally exploit this mapping approach, an SRAM CIM–aware pruning algorithm committed to generating kernels containing 16 groups of weights that are all zero is also proposed. In Section IV, how this algorithm is designed to further improve performance is described in detail.

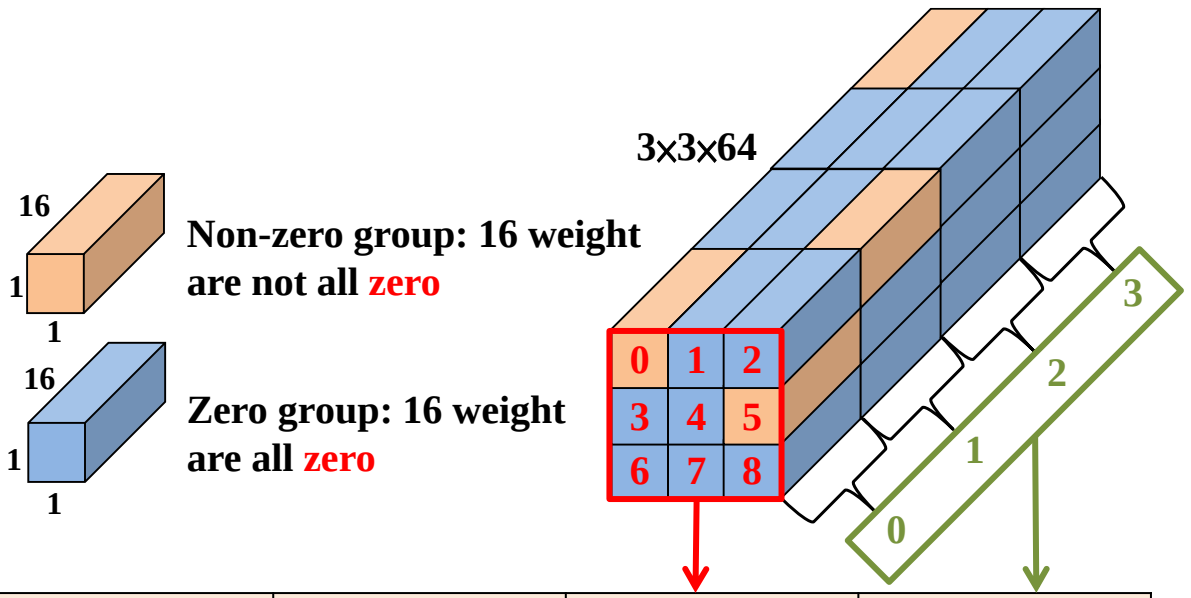

| 1st group ? | Total group-1 | 3X3 position | Channel position |
|---|---|---|---|
| Index[15] | Index[14:9] | Index[8:5] | Index[4:0] |
| 1 | 6'b000011 | 0 | 0 |
| 0 | 6'b000011 | 0 | 3 |
| 0 | 6'b000011 | 2 | 1 |
| 0 | 6'b000011 | 5 | 0 |

Fig. 6 Bit [15] records whether a given group is the first group of the kernel. Bit [14:9] records the total number of non-zero groups contained in the kernel, Bit [8:5] records the position of the group in the 3 × 3 kernel order, and Bit [4:0] is recorded as the position of the channel order direction.

3) Index Code Compression

Because the aforementioned proposed weight sparsity mapping method is used, the weight-groups stored in the SRAM CIM are no longer in order. Different kernels with the same size could cause different memory usage and sequences because the number of zeros of each kernel differs. Therefore, the system requires an index code to identify which position of the origin kernel each weight-group belongs to. The 16 weight-groups in a group-set can be remapped back to the same position of the original 16 kernels; therefore, each group-set in the SRAM CIM can be represented with one index code. A 16-bit index code is designed to contain information regarding the 16 nonzero weight-groups in the SRAM CIM to enable the SAS system to obtain the corresponding position of the IFM data. Fig. 6 illustrates an example of the index code recording the location of the 16 nonzero weight-groups of kernels. With this method, only nonzero weights and their index codes are stored, consuming up to 75 times less memory than storing all weights in the kernels.

### C. Four-Core Multicycle System

The maximum operating frequency of the CIM macro can reach approximately 100 MHz. Four CIM cores are used in this work; these cores require access to the IFM SRAM and OFM SRAM during calculations. However, both single and dual port SRAM fail with four concurrent access requests. Therefore, a multicore shunter is designed to address this concern. Initially, the cores' operating frequency is restricted to 100 MHz because of the computing effort of the CIM macro and the accumulators. Since most tasks in the top-level system, except the CIM core, are mainly straightforward data processing and memory accessing, achieving a high frequency is feasible if the system is designed appropriately. Taking the shunter as the demarcation point, the operating frequency of the top-level system is raised to exactly four times that of the cores.

With the system (top level) and the CIM cores' operating frequency at 400 MHz and 100 MHz, respectively, this architecture circumvents the limitations the CIM macro could place on overall system performance. The shunter distributes the IFM and OFM SRAM access requests from the four CIM cores equally. As illustrated in Fig. 7, CIM core 1's request for IFM data is sent to the IFM SRAM in the first cycle, and CIM core 2's request is sent in the second cycle (and so forth). Once the shunter sends one core's access request to the FM SRAM, four system cycles are required for the same core to access the FM SRAM again, and the CIM core also has sufficient time to perform complicated calculations during the four cycles.

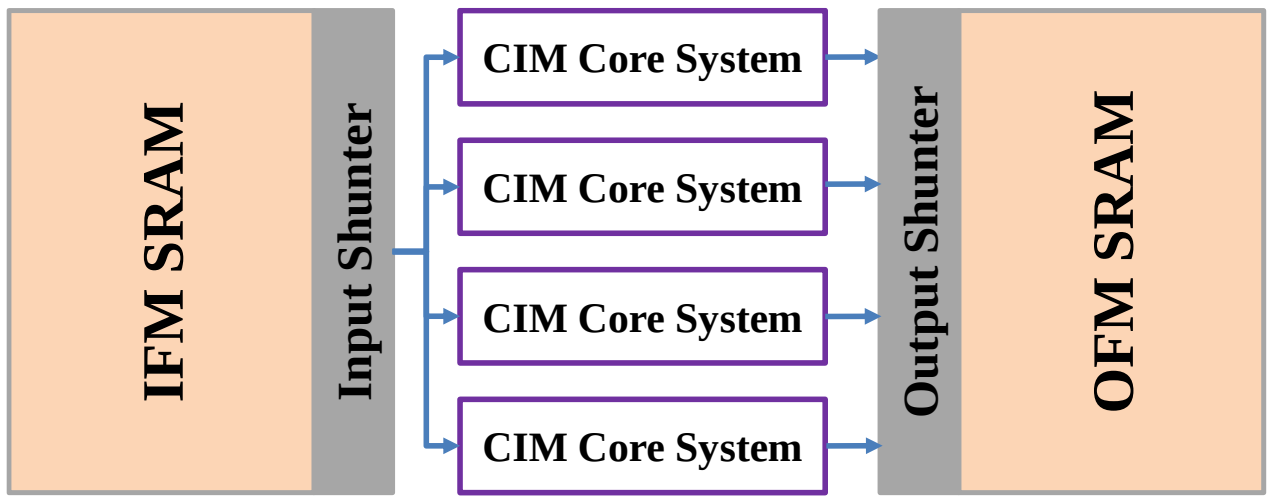

| System Cycle | 1 | 2 | 3 | 4 | 5 | 6 |
|---|---|---|---|---|---|---|
| Core1 | Request! | Calculation | Calculation | Calculation | Request! | Calculation |
| Core2 | Request! | Wait | Calculation | Calculation | Calculation | Request! |
| Core3 | Request! | Wait | Wait | Calculation | Calculation | Calculation |
| Core4 | Request! | Wait | Wait | Wait | Calculation | Calculation |
| Shunter accept | Core1 | Core2 | Core3 | Core4 | Core1 | Core 2 |

Fig. 7 Four-core multicycle system and shunter block of MARS

## IV. SRAM-Aware Model Compression Algorithm

### A. CIM-Aware Pruning

The hardware architecture of MARS, including how to address the calculation of 0, is introduced in section III. In this section, a sparse algorithm that considers the limitations of the SRAM CIM macro to further optimize the design of MARS, is proposed.

1) Group Lasso Regularization

The main objective of pruning is to obtain a sparse neural network that has considerably fewer parameters than the original dense network. The training objective function for a DNN for classification is given as follows:

$$E(w) = L(w) + \frac{\lambda}{2}R(w) + \frac{\lambda_g}{2}\sum_{l=1}^{L} R_g(w^l) \qquad (1)$$

where w represents the collection of all weights in the CNN, $L(w)$ is the loss function, $R(\cdot)$ is non-structured regularization on every weight. $R_g$ is the structure sparsity regularization on each layer, and $\lambda$ and $\lambda_g$ are the hyperparameters to control the trade-off between classification accuracy and the sparsity ratio. In [21], group lasso is adopted for $R_g$ to effectively minimize the number of parameters in groups to near zero.

2) Proposed CIM-aware pruning

Inspired by [21], a structured sparsification method is proposed to train the networks and distribute the zero-value weight according to the SRAM CIM macro structure. Fig. 8 illustrates the concept of the proposed SRAM-aware pruning method. Because of the mapping constraint of the SRAM CIM macro, the positions of zeros on different kernels can only be in the same place. In addition, because of hardware limitations, each input pixel can only be multiplied with eight weights, corresponding to the weight from the same position of eight different kernels (Fig. 8, marked in red). In the hardware, these eight weights must all have a value of zero at the same time to be skipped, otherwise erroneous calculations will take place. Therefore, if the CNN model has more of such distributions of zero, the SRAM CIM can achieve more efficient acceleration. The regularization of the proposed group regularization function can be expressed as follows:

$$E(w) = L(w) + \frac{\lambda}{2}R(w) + \frac{\lambda_g}{2}\sum_{l=1}^{L} R_{gsw}(w^l) \qquad (2)$$

where $R_{gsw}$ is the regularization of the group lasso on a group of weights that mapping on a single column of SRAM CIM:

$$R_{gsw}(w^l) = \sum_{c_l=1}^{C_l}\sum_{m_l=1}^{M_l}\sum_{k_l=1}^{K_l}\sum_{f_l=1}^{F_l/\alpha}\left\|W^l_{\alpha(f_l-1)+1:\alpha(f_l-1)+\alpha,c_l,m_l,k_l}\right\|_g \qquad (3)$$

where $||\bullet||_g$ denotes the group lasso; $l$ is the number of layers; $C_l$ is the number of channels in one layer; $M_l$ and $K_l$ are the spatial height and width, respectively; $F_l$ is the total filter number of a layer; $\alpha$ is the number of weights that can be calculated in a single cycle for each pixel. By adding this regularization term after the loss function, the weight in a predefined structure with a small accuracy loss can be pruned.

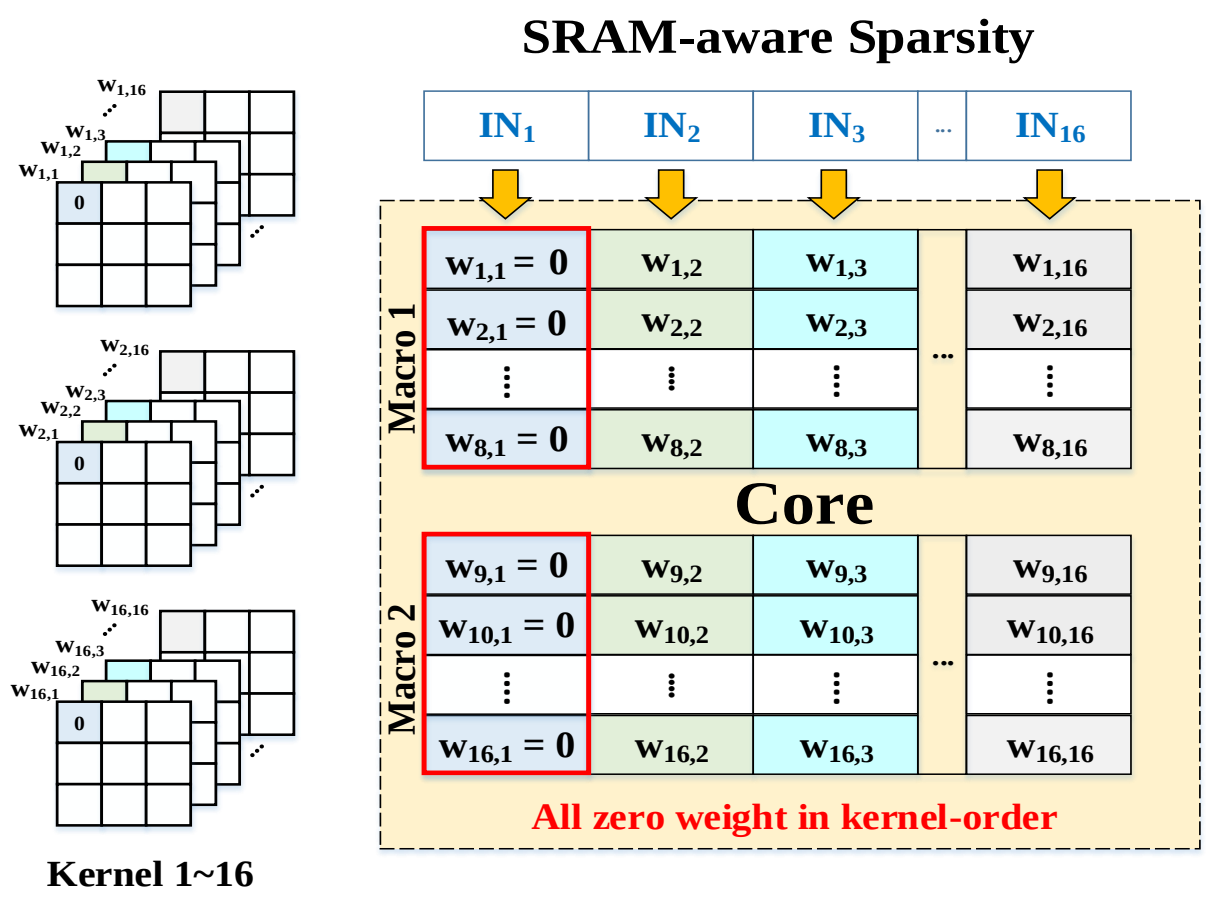

Fig. 8 SRAM-aware sparsity on two SRAM-CIM cores. For each input pixel, each core can calculate 8 weights and calculate 16 input pixels in parallel in one cycle. To deploy the sparse CNN efficiently, arranging the positions of zeros in different kernels in the same place is ideal.

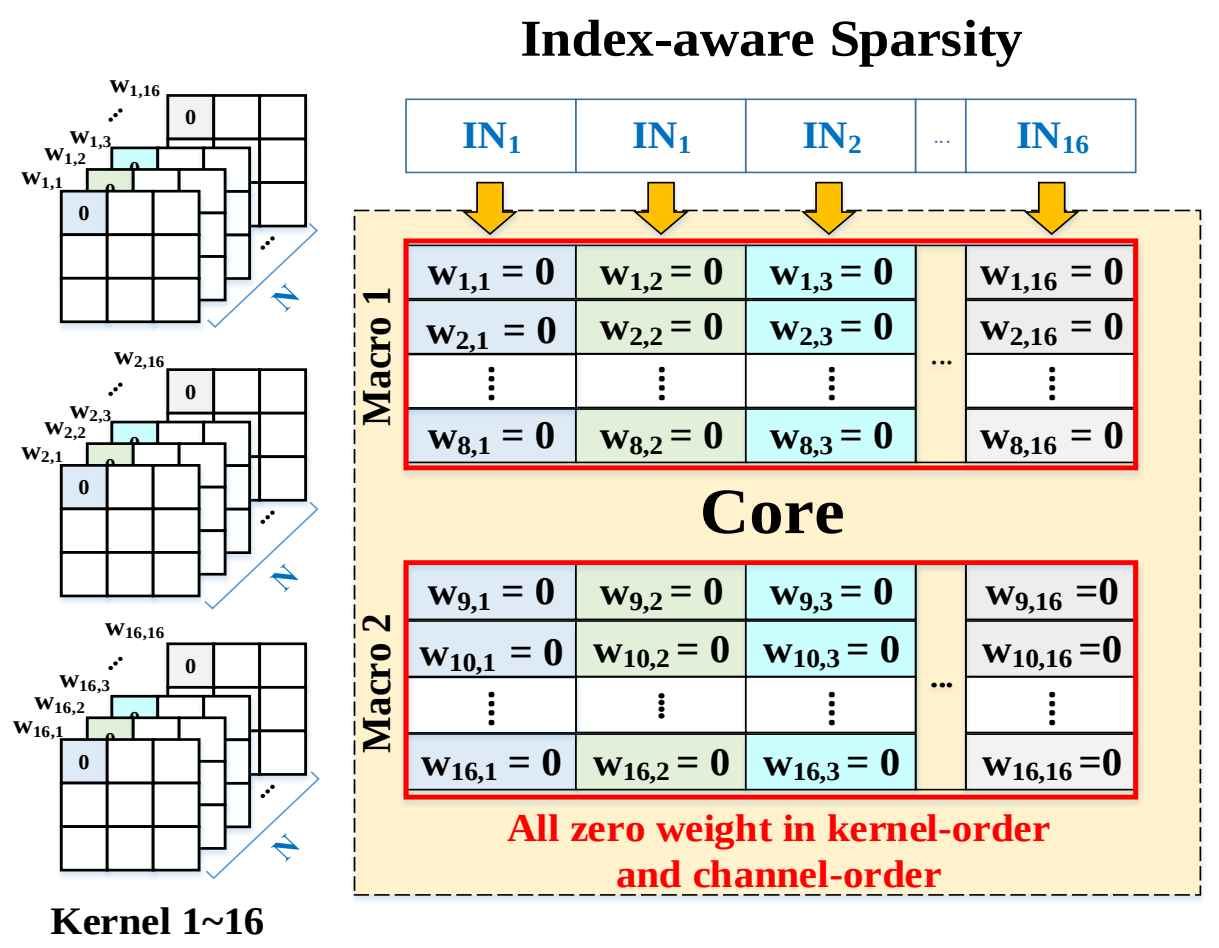

Fig. 9 Index-aware sparsity enables the whole core to share the same index.

## B. *Index-Aware Pruning*

In section A, group regularization is exploited to sparsify the weights in the same positions of every α different kernel and then perform pruning based on the structure of the former step. Using this method, the accuracy loss can be reduced to a negligible level. Even if the pruning ratio is high, the accuracy can be restored to levels near that of the original model through retraining. To denote the location of the remaining weights, using 1 or 0 to store the surviving weight and pruned weight is a straightforward method [38]. Generally the index storage requirement is associated with the grain size. Therefore, filter-wise pruning and channel-wise pruning have less index storage overheads than fine-grained sparsity does. Despite fine-grained pruning having a higher compression rate, the index storage concern cannot be ignored. Moreover, the index storage requirement is associated with energy consumption. The smaller the register fetched from the memory, the more transmission energy can be saved.

To mitigate the problem of index storage, thes zero position in different channels on the same kernel needs to be the same. Fig. 9 describes this method in detail. If the zero position on N channels in every α kernel is the same, the index storage requirement can be reduced by N× to effectively achieve the aforementioned goal. The group regularization term is rewritten from equation (3) as follows:

$$R_{gsw}(w^l) = \sum_{c_l=1}^{C_l}\sum_{m_l=1}^{M_l}\sum_{k_l=1}^{K_l/N}\sum_{f_l=1}^{F_l/\alpha}\left\|W^l_{\alpha(f_l-1)+1:\alpha(f_l-1)+\alpha,N(c_l-1)+1:N(c_l-1)+N,m_l,k_l}\right\|_g \qquad (4)$$

where $N$ is a hyper-parameter that denotes N weights in the direction of the channel as a group; this ensures that the model simultaneously learns the sparsity of the weights on the same position in different N channels. The weights that can be pruned will have more regularity, thus saving more index storage costs.

## C. *Quantization Algorithm*

### 1) Activation Quantization

A straight-through estimator (STE) [40] is applied on the input activations of each layer. To facilitate activation quantization during hardware implementation, the clip function is used to ensure the input falls within [0, 1] instead of normalization. The activation quantization function is as follows:

$$A_l^q = Q_A(A_l, 2^{b_A}) = \frac{round\left(clamp(A_l,0,1)\times(2^{b_A}-1)\right)}{(2^{b_A})} \qquad (5)$$

where $A_l$ denotes the activations (or inputs) of the $l$-th layer in the CNN and $b_A$ denotes the number of bits of activation.

Table I
Comparison with prior arts

| | ISSCC19 [40] | ISSCC19 [41] | ISSCC20 [39] | | **MARS** *2 | | | |
|---|---|---|---|---|---|---|---|---|
| CIM Technology | 55nm | 28nm | 65nm | | **28nm** | | | |
| CIM frequency (MHz) | 98~320 | 400 | 50~100 | | **100** | | | |
| Area ($mm^2$) | 0.037 (macro) | 0.22 (macro) | 9.00 (system) | | **6.83 (system)** | | | |
| Single CIM macro power (mW) | 0.02~0.07 | N/A | 1.7~3.8 | | **1.9~2.7** | | | |
| Weight precision | 2/5 | 1 | 4/8 | | **4/8** | | | |
| Activation precision | 1/2/4 | 1 | 2/4/6/8 | | **4/8** | | | |
| Sparsity support | N/A | N/A | activation/weight sparsity | | **weight sparsity** | | | |
| Test Network | ResNet20 | AlexNet | VGG16 | ResNet18 | **VGG16** | | **ResNet18** | |
| Dataset | CIFAR10 | MNIST | CIFAR10 | | **CIFAR10** | **CIFAR100** | **CIFAR10** | **CIFAR100** |
| Frame per second (FPS) | @w2a1 *1 | @w1a1 *1 | @w8a4 *1 | | **@w8a4** *1 | | | |
| | N/A | N/A | 268 | 182 | **714** | **483** | **711** | **403** |
| | | | | | **@w8a8** | | | |
| | | | | | **540** | **377** | **403** | **339** |
| Avg. throughput (GOPs) | 56.24 | N/A | 22.30~26.83 | 49.86~59.96 | **@w8a4** | | | |
| | | | | | **445** | **301** | **778** | **441** |
| | | | | | **@w8a8** | | | |
| | | | | | **336** | **235** | **441** | **371** |
| CIM macro Energy efficiency (TOPs/W) | 72.1 | 119.7 | 13.1 | 15.8 | **@w8a4** | | | |
| | | | | | **52.3** | **49.8** | **88.2** | **60.6** |
| | | | | | **@w8a8** | | | |
| | | | | | **29.7** | **23.8** | **37.6** | **38.0** |

*1: @wnam means this model uses n-bit weight and m-bit activation.
*2: The throughput and energy efficiency of MARS are estimated value, and the power and operating frequency of SRAM CIM information are referred from [18]. The area of MARS is estimated without the IO pad.

2) Weight Quantization with BN Fusion

Unlike in previous methods [41] in which the BN is first fused into the weights and then the weights are quantized, in this work, the weights proceeded through some processes before fusing with the BN.

The complete algorithm can be divided into three parts. First, the weights are split into several groups and the number of groups, G, is determined by the numbers of BLs that can be turned on in a single cycle. Second, the value range of each group of weights is limited to [-1, 1] by tanh function as follows:

$$\widehat{W}_{l,g} = \frac{tanh(W_{l,g})}{max(|tanh(W_{l,g})|)} \quad (6)$$

where $W_{l,g}$ denotes the weight matrix of $g$-th groups in the $l$-th layer in the CNN.

Third, BN is fused into the processed weights $\widehat{W}$, and the clamp function is used to ensure that the value range is limited to [-1, 1]. The full equation is as follows:

$$\overline{W}_{l,k} = \text{clamp}\left(\frac{\gamma \times \widehat{W}_{l,k}}{\sqrt{\sigma^2_{A_{l,ch}} + \epsilon}}, -1, 1\right) \quad (7)$$

where $\widehat{W}_{l,k}$ denotes the $k$-th kernel in the $l$-th layer in the CNN, $\gamma$ is a trainable parameter that is the same as the BN parameter, $\sigma^2_{A_{l,ch}}$ is the mini-batch variance that is updated by exponential moving average, and $\epsilon$ is a very small value to avoid the denominator being zero.

The last part involves using symmetric quantization and applying STE to the weights. The equation is as follows:

$$W^q_{l,k} = Q_W(\overline{W}_{l,k}, 2^{b_W}) = \frac{round\left(\overline{W}_{l,ch} \times (2^{b_W - 1} - 1)\right)}{(2^{b_W - 1})} \quad (8)$$

where $b_W$ denotes the number of bits of weight. For $b_W = 4$, the final value of weights will be in [-7, -6, …, 0, …, 6, 7]. Using this approach, the values can be implemented directly in the hardware.

## V. Experimental Results

This section is divided into two parts. The first part summarizes the performance of the proposed SRAM CIM-based accelerator, and the second part presents the results of the proposed model compression algorithm on VGG16 [1] and ResNet18 [2] with CIFAR10 and CIFAR100.

### *A. SRAM CIM–Based Accelerator Performance*

1) Performance of the MARS architecture

The simulation results of MARS are compared with the accelerator without supporting sparsity circuit (baseline). The baseline adopted the same architecture as MARS without skipping the calculation and storing of zero weight. In the experiment, the activation and weight precisions were both 4/8-bits on VGG-16 and ResNet-18 using CIFAR-10 and CIFAR-100 datasets at 400MHz system operating frequency; the SRAM CIM macro was operating at 100 MHz. Fig. 10 depicts that MARS enhanced the performance for different datasets compared to the baseline. MARS achieved a performance at most 13 times better than the baseline in VGG16 on CIFAR10. According to the measurement result of [18], from which the SRAM CIM MARS adopted, the evaluation of average macro energy efficiency of the whole network reached at most 88 TOPs/W with MARS in ResNet18 on CIFAR10. Fig. 11 shows the amount of feature map memory (SRAM1,2) access reduction of MARS comparing to the baseline. As the network layer went deeper, the compression rate was greater, and the gap between the baseline and MARS to request the amount of data from the memory was larger. In VGG16 and ResNet18 convolutional layer, MARS reduced at most 290 times and 440 times the amount of memory access than the baseline, respectively.

2) Comparison with prior arts

Table I shows the comparison result between MARS and state-of-the-art SRAM CIM macro [42], [43] and architecture [44]. In this table, the power efficiency were estimated by refering the measurement result of [18] whose power was about 1.9~2.7mW at 100 MHz operating frequency. MARS's macro peak energy efficiency could achieve 694 TOPs/W, which is 5.8 times higher than state-of-the-art [44]. These improvements mainly result from omitting the zeros and saving memory storage to reduce data transfer by using the sparsity-aware mechanism.

## B. *Model Compression Algorithm*

1) Experimental Setup

All experiments were trained from scratch, implemented with Pytorch [45], and performed on NVIDIA V100. VGG16 and ResNet18 were trained with a batch size of 128 for 400 epochs on both CIFAR10 and CIFAR100 and optimized by using stochastic gradient descent (SGD). The learning rate was initially set at 0.05 and divided by 10 for every 100 epochs. For the hyper-parameters in the sparsity algorithm, α and N were set to 16.

2) Results for CIM-aware Pruning

In this section, the pruning result on convolution layers is discussed because convolution layers perform most of the multiply-accumulate operation (MAC) in the network calculations. To deploy the hardware, in addition to the floating point, the sparse model was quantized to different precisions according to the specification of the CIM macro by using the quantization algorithm proposed in Section IV. C. The zero-rows proportion was used to measure the effect of the CIM-aware pruning, because the zero-row proportion represented the number of rows that could be skipped without being stored in the SRAM CIM during actual hardware operation. Table II summarizes the performance of the proposed CIM-aware pruning algorithm on CIFAR10 and CIFAR100, respectively. Because of hardware limitations, the values of α and N were set to 16 to minimize the index storage requirement and fulfill the calculate limitation of CIM macro.

In the CIFAR10 experiments, regardless of whether VGG16 or ResNet18 were full-precision or quantized-sparse models, similar compression rates were achieved, and the error rate did not increase by more than 0.9% compared with that of the original model in most cases. Moreover, the sparse model achieved a 100×compression rate when the precision of the weight and activation was 4 bits. For the CIFAR100 data set, the compression rate of the quantized model and the full-precision model was slightly different because of the increased complexity of data classification. A considerable compression rate was still achieved for VGG16 and most cases of ResNet18 without much accuracy loss.

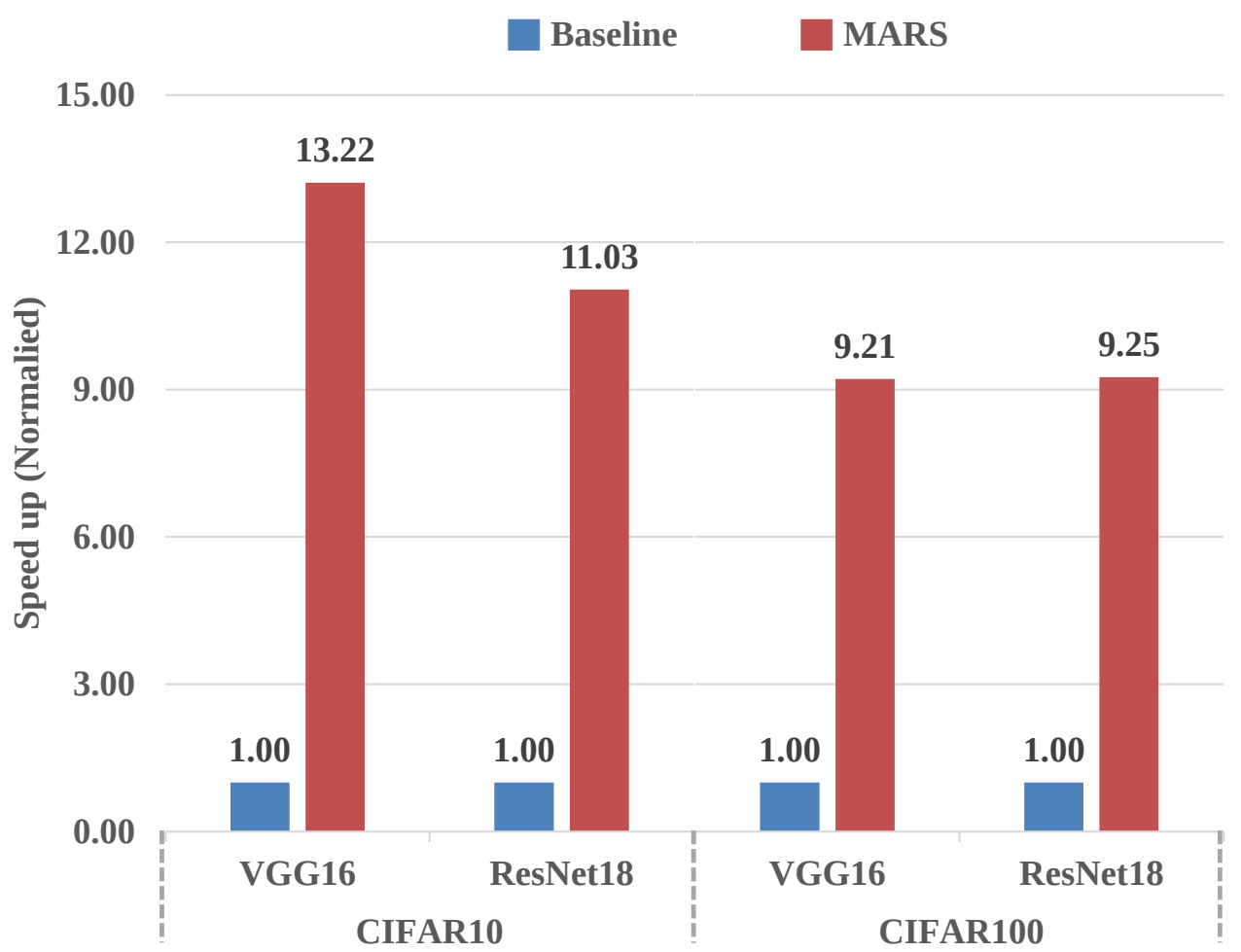

Fig. 10 Normalized performance speedup of MARS comparing to the baseline. MARS enhanced the performance of processing an image. Both baseline and MARS performance evaluation include the time of loading weight to the CIM macro.

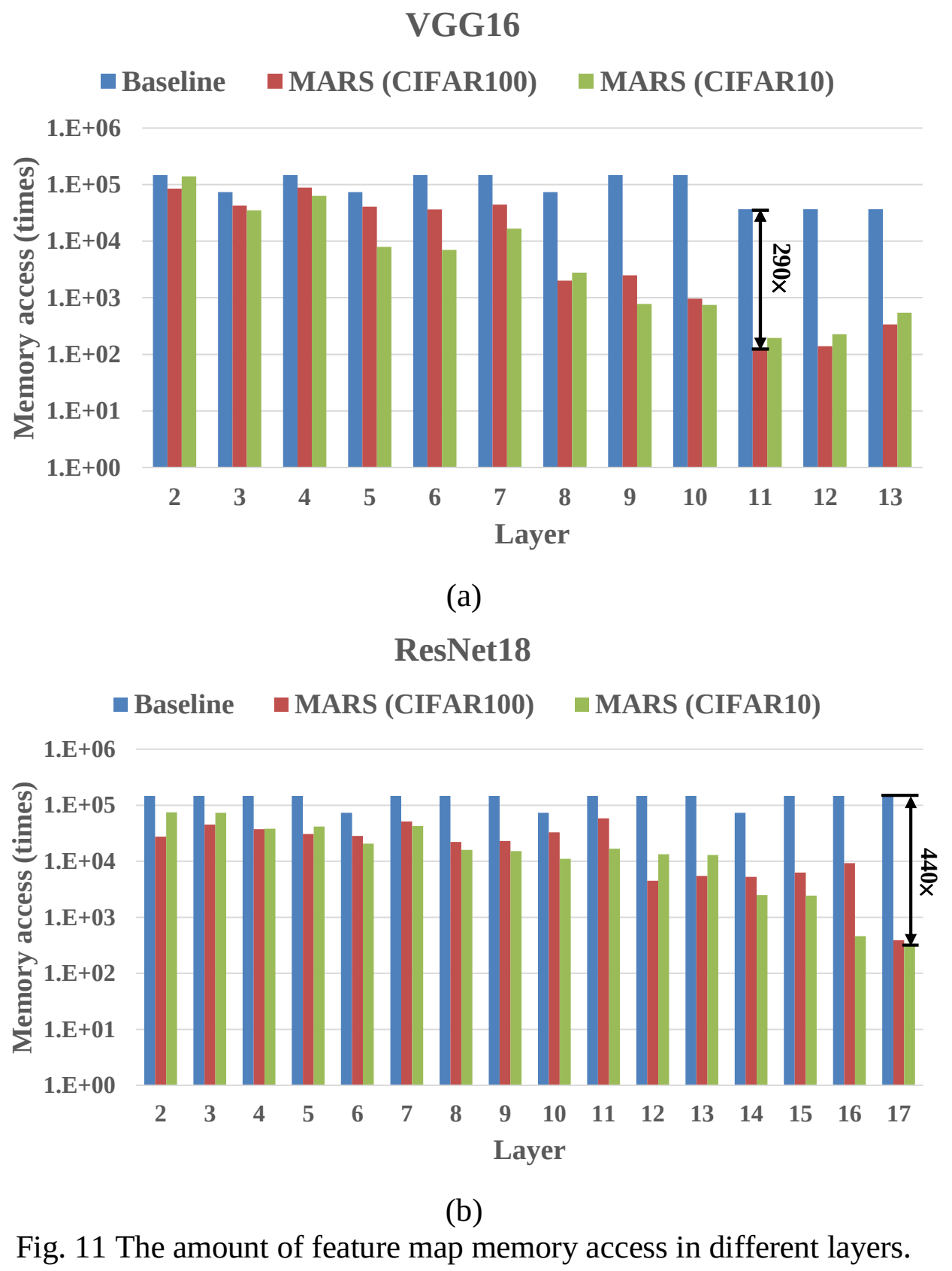

Fig. 11 The amount of feature map memory access in different layers.

Table II
Results of combining index-aware pruning and quantization algorithm on CIFAR10 and CIFAR100

| Model | Bit-Width (W/A) | Original Accuracy (%) | Sparsity(%) | Sparsity Accuracy (%) | Compression Rate |
|---|---|---|---|---|---|
| **CIFAR10** | | | | | |
| VGG16 | 32/32 | 92.83 | 97.00 | 92.71 | 33.3x |
| | 8/8 | 92.26 | 96.00 | 92.53 | 100x |
| | 8/4 | 92.60 | 95.00 | 92.32 | 80x |
| | 4/4 | 91.80 | 95.00 | 92.44 | 160x |
| ResNet18 | 32/32 | 93.10 | 95.00 | 92.84 | 20x |
| | 8/8 | 94.04 | 95.00 | 92.80 | 80x |
| | 8/4 | 93.30 | 95.00 | 92.74 | 80x |
| | 4/4 | 93.30 | 94.00 | 92.60 | 133.3x |
| **CIFAR100** | | | | | |
| VGG16 | 32/32 | 69.53 | 93.00 | 68.71 | 14.6x |
| | 8/8 | 70.52 | 91.00 | 71.62 | 44.4x |
| | 8/4 | 70.08 | 91.00 | 70.62 | 44.4x |
| | 4/4 | 71.04 | 88.00 | 70.82 | 66.7x |
| ResNet18 | 32/32 | 71.22 | 92.00 | 69.30 | 12.5x |
| | 8/8 | 72.99 | 85.00 | 71.24 | 26.7x |
| | 8/4 | 71.32 | 85.00 | 71.84 | 26.7x |
| | 4/4 | 71.09 | 85.00 | 70.82 | 53.3x |

Table III
Comparison of quantized models accuracy on CIFAR10 and CIFAR100

| Model | Method | Bit-width W/A | Accuracy (%) CIFAR10 | Accuracy (%) CIFAR100 |
|---|---|---|---|---|
| VGG16 | DoReFa | 32/32 | 92.83 | 69.53 |
| | | 8/8 | 92.98 | 69.76 |
| | | 8/4 | 92.86 | 70.96 |
| | | 4/4 | 92.68 | 69.19 |
| | This work | 8/8 | 93.14 | 71.88 |
| | | 8/4 | 92.74 | 70.81 |
| | | 4/4 | 92.82 | 70.91 |
| ResNet18 | DoReFa | 32/32 | 93.10 | 71.22 |
| | | 8/8 | 93.14 | 71.57 |
| | | 8/4 | 93.42 | 71.15 |
| | | 4/4 | 93.20 | 69.97 |
| | This work | 8/8 | 93.89 | 72.99 |
| | | 8/4 | 93.83 | 71.49 |
| | | 4/4 | 93.18 | 70.83 |

This model was trained without BN.

3) Comparison of Quantization Algorithm

Table III presents the comparison between the Dorefa [25] quantization algorithm without a BN network and the proposed quantization algorithm (not trained with the proposed sparsity algorithm). Under the condition with the same bit-width of weight and activation, the accuracy of the proposed method had a better performance, especially for more complex datasets. For the 4/4 case on CIFAR100 for VGG16 and ResNet18, the accuracy of VGG16 was increased by 0.98%, and the accuracy of ResNet18 only decreased by 0.17%, compared to the baseline.

4) Index Storage Analysis

In this experiment, different N used when training under equation (4) is discussed. The baseline in this experiment was N = 1, referring to the case that did not consider index saving. Fig. 12 illustrates the result of VGG16 and ResNet18 training on CIFAR10. α was set to 16; N was set to 1, 4, 8, 16, and 32, respectively. The compression rate decreased when the value of N increased; a loss of only 1% in sparsity ratio was observed when N was 16; the index storage requirement was saved 16 times. Moreover, no sparsity ratio loss was observed in ResNet18 between N of 4 and N of 16, and only a 0.5% loss in sparsity ratio was observed compared with baseline. This experiment indicates that training under equation (4) is an efficient method to minimize the index storage requirement while not causing accuracy loss or much sparsity ratio loss. However, a considerable decrease in accuracy and sparsity ratio was observed when the value of N was 32. The sparsity ratio decreased by over 2% on both CIFAR10 and CIFAR100. Therefore, N was set to 16 to prune the model and implement the hardware.

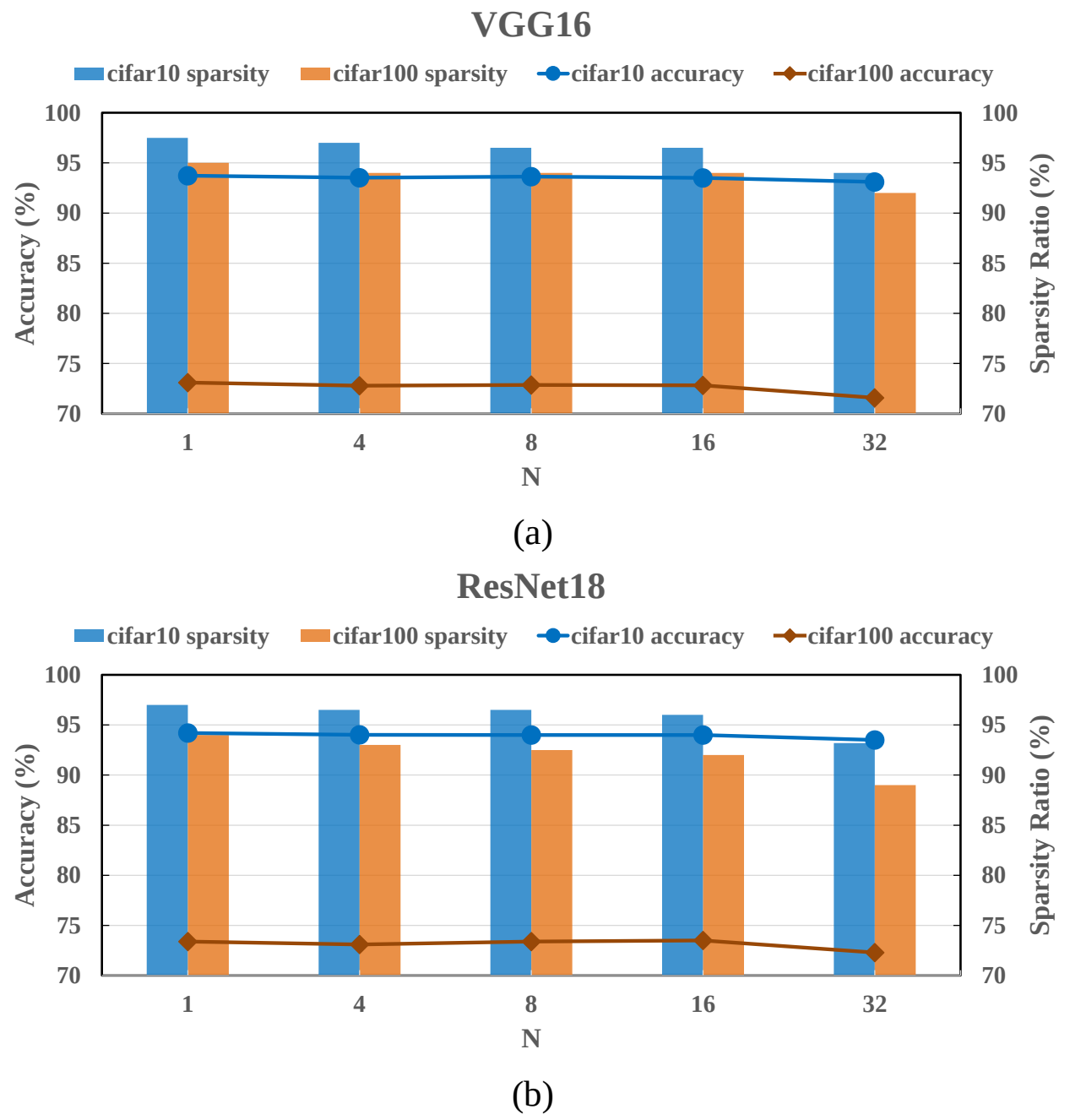

Fig. 12 The accuracy (line) and compression rate (bar) respect to different *N*. CIFAR10 is blue and CIFAR100 is red. (a) is VGG16 and (b) is ResNet18.

Table IV shows the memory compression rate in different layers of VGG16 on CIFAR10. All weights in the networks were quantized in 8-bit. Therefore, for the case of $3 \times 3 \times 512 \times 512$, 18 Mb of SRAM was originally required to store

Table IV
Compression of memory size. Original memory storage refers to storage of all the weights. The weights in both storage methods were quantized to 8 bits.

| Different layer | Original Memory Storage (Mb) | Memory storage with our method (Kb) | | | Memory size Compression |
|---|---|---|---|---|---|
| | | C.R. | Index | Weight | |
| 3x3x64x64 | 0.28 | 5% | 2.14 | 273.60 | 1.04x |
| 3x3x64x128 | 0.56 | 50% | 2.25 | 288.00 | 1.98x |
| 3x3x128x128 | 1.13 | 56.6% | 3.91 | 488.97 | 2.29x |
| 3x3x128x256 | 2.25 | 61.6% | 6.91 | 884.74 | 2.58x |
| 3x3x256x256 | 4.50 | 93.2% | 2.46 | 313.34 | 14.59x |
| 3x3x256x512 | 9.00 | 97.8% | 1.58 | 202.75 | 45.10x |
| 3x3x512x512 | 18.00 | 98.7% | 1.87 | 239.62 | 73.33x |

*C.R. is the compression rate

all the weights. However, with the proposed method, only 239.62 Kb and 1.87 Kb were required to store the weights and index, respectively.

## VI. Conclusion

In this paper, a software and hardware co-design approach is proposed to design an SRAM CIM-based CNN accelerator and an SRAM CIM-aware model compression algorithm. The proposed SRAM CIM-based CNN accelerator used eight SRAM CIM macros as PEs and supported sparse CNN computing. With this proposed architecture, the overall computing efficiency was increased by 13 times. The proposed sparsity algorithm considering the CIM architecture achieved a high compression ratio. When both weight and activation were floating points, the sparsity rate was 0.97 for VGG16 on CIFAR10. The proposed quantization algorithm could fuse BN into the algorithm to lessen the high-precision MAC required by BN. For the case of 4-bit weight and 4-bit activation on CIFAR100, the accuracy of the proposed quantization method increased by 0.98% for VGG16 and only decreased by 0.17% for ResNet18, compared to the baseline. Furthermore, the index storage method could more effectively decrease the number of bits required to store the positions of non-zero weights.

## Acknowledgment

We are thankful to the Taiwan Computing Cloud (TWCC) for computer time and facilities.

## References

[1] K. Simonyan and A. Zisserman, "Very deep convolutional networks for large-scale image recognition", in ICLR, 2015.
[2] K. He, et al., "Deep Residual Learning for Image Recognition", in CVPR, 2016.
[3] S. Ioffe, et al., "Batch Normalization: Accelerating Deep Network Training by Reducing Internal Covariate Shift", in ICML, 2015.
[4] A. G. Howard, et al., "MobileNets: Efficient Convolutional Neural Networks for Mobile Vision Applications", in arXiv:1704.04861, 2017.
[5] Y.-H. Chen, et al., "Eyeriss: An Energy-Efficient Reconfigurable Accelerator for Deep Convolutional Neural Networks", IEEE Journal of Solid State Circuits (JSSC), ISSCC Special Issue, Vol. 52, No. 1, pp. 127-138, 2017.
[6] K. Ueyoshi, et al. "QUEST: A 7.49 TOPS multi-purpose log-quantized DNN inference engine stacked on 96MB 3D SRAM using inductive-coupling technology in 40nm CMOS", in ISSCC, 2018.
[7] Y. Zhe, et al., "Sticker: A 0.41-62.1 TOPS/W 8Bit neural network processor with multi-sparsity compatible convolution arrays and online tuning acceleration for fully connected layers", in VLSI, 2018.
[8] X. Xu, et al., "Scaling for edge inference of deep neural networks". Nature Electronics, 2018.
[9] C.-X. Xue, et al., "A 1Mb Multibit ReRAM Computing-In-Memory Macro with 14.6ns Parallel MAC Computing Time for CNN Based AI Edge Processors", in ISSCC, 2019.
[10] J. Zhang, et al., "In-memory computation of a machine-learning classifier in a standard 6T SRAM array", IEEE J. Solid-State Circuits, vol. 52, no. 4, pp. 915–924, 2017.
[11] S. Xin et al., "A Dual-Split 6T SRAM-Based Computing-in-Memory Unit-Macro With Fully Parallel Product-Sum Operation for Binarized DNN Edge Processors", IEEE Transactions on Circuits and Systems I: Regular Papers. PP. 1-14. 10.1109/TCSI.2019.2928043. 2019.
[12] A. Biswas et al., "Conv-RAM: An energy-efficient SRAM with embedded convolution computation for low-power CNN-based machine learning applications," in ISSCC, 2018.
[13] Z. Jiang, et al., "XNOR-SRAM: In-memory computing SRAM macro for binary/ternary deep neural networks", in VLSI, 2018.
[14] D. Qing et al.,"A 0.3V VDDmin 4+2T SRAM for searching and in-memory computing using 55nm DDC technology", in VLSI, 2017.
[15] A. Amogh et al., "Xcel-RAM: Accelerating Binary Neural Networks in High-Throughput SRAM Compute Arrays", IEEE Transactions on Circuits and Systems I: Regular Papers 66: 3064-3076, 2018.
[16] V. Hossein et al., "A 64-Tile 2.4-Mb In-Memory-Computing CNN Accelerator Employing Charge-Domain Compute", IEEE Journal of Solid-State Circuits. PP. 1-11. 10.1109/JSSC.2019.2899730, 2019.
[17] X. Si, et al., "A Twin-8T SRAM Computation-in-Memory Unit-Macro for Multibit CNN-Based AI Edge Processors", IEEE J. of Solid-State Circuits, vol.55, no. 1, pp. 182–202, 2019.
[18] X., Si, et al., "A 28nm 64Kb 6T SRAM Computing-in-Memory Macro with 8b MAC Operation for AI Edge Chips", in ISSCC, 2020.
[19] H. Li, et al., "Pruning Filters For Efficient Convnets", in ICLR, 2017.
[20] Z. Liu, et al., "Learning Efficient Convolutional Networks through Network Slimming", in ICCV, 2017.
[21] W. Wen, et al, "Learning Structured Sparsity in Deep Neural Networks", in NIPS, 2016.
[22] S. Han, et al., "Learning both Weights and Connections for Efficient Neural Networks", in NIPS, 2015.
[23] I. Hubara, et al., "Binarized Neural Networks", in NIPS, 2016.
[24] C. Zhu, et al., "Trained Ternary Quantization", in ICLR, 2017.
[25] S. Zhou, et al., "Dorefa-Net: Training Low Bitwidth Convolutional Neural Networks With Low Bitwidth Gradients", in CVPR, 2016.
[26] J. Choi, et al., "PACT: Parameterized Clipping Activation for Quantized Neural Networks", in arXiv:1805.06085, 2018.
[27] S. Wu, et al., "Training and inference with integers in deep neural networks", In ICLR, 2018.
[28] Z. Cai, et al., "Deep learning with low precision by half-wave Gaussian Quantization", in CVPR, 2017.
[29] D. Miyashita, et al., "Convolutional neural networks using logarithmic data representation", in arXiv:1603.01025, 2016.
[30] C. Baskin, E. Schwartz, et al., "Uniq: Uniform noise injection for non-uniform quantization of neural networks", in arXiv:1804.10969, 2018.
[31] A. Shafiee, et al. "ISAAC: A Convolutional Neural Network Accelerator with In-Situ Analog Arithmetic in Crossbars", in ISCA, 2016.
[32] H. Ji et al., "ReCom: An Efficient Resistive Accelerator for Compressed Deep Neural Networks", in DATE, 2018.
[33] J. Lin, et al. "Learning the Sparsity for ReRAM:Mapping and Pruning Sparse Neural Network for ReRAM Based Accelerator" in ASPDAC, 2019.
[34] A. Parashar, et al., "SCNN: An accelerator for compressed-sparse convolutional neural networks", in ISCA, 2017.
[35] H. Song, et al. "EIE: Efficient Inference Engine on Compressed Deep Neural Network", ACM SIGARCH Computer Architecture News. 44. 10.1145/3007787.3001163, 2016.
[36] H. Song, et al. "Deep Compression: Compressing Deep Neural Networks with Pruning, Trained Quantization and Huffman Coding", in ICLR, 2016.
[37] T. Frederick, and M. Greg. "CLIP-Q: Deep Network Compression Learning by In-parallel Pruning-Quantization", in CVPR, 2018.
[38] A. Alessandro et al., "NullHop: A Flexible Convolutional Neural Network Accelerator Based on Sparse Representations of Feature Maps," IEEE

Transactions on Neural Networks and Learning Systems, vol. 30, no. 3, pp. 644–656, 2019.

[39] H. Mao, et al., "Exploring the regularity of sparse structure in convolutional neural networks". Workshop paper in CVPR, 2017

[40] Y. Bengio, et al., "Estimating or propagating gradients through stochastic neurons for conditional computation", In arXiv:1308.3432, 2013.

[41] B. Jacob, et al., "Quantization and Training of Neural Networks for Efficient Integer-Arithmetic-Only Inference" in CVPR 2018.

[42] J. Yue, et al., "A 65nm Computing-in-Memory Based CNN Processor with 2.9-35.8TOP/W System Energy Efficiency Using Dynamic Sparsity Performance Scaling Architecture and Energy Efficient Inter/Intra Macro Data Reuse", in ISSCC, 2020.

[43] X. Si et al., "A Twin-8T SRAM Computation-In-Memory Macro for Multiple-Bit CNN-Based Machine Learning", in ISSCC, 2019.

[44] J. Yang et al., "Sandwich-RAM: An Energy-Efficient In-Memory BWN Architecture with Pulse-Width Modulation", in ISSCC, 2019.

[45] P. Adam, et al., "PyTorch: An Imperative Style, High-Performance Deep Learning Library", 2019.